\documentclass[aps,prl,twocolumn,groupedaddress,showpacs]{revtex4}
\usepackage{graphicx}
\usepackage{hyperref}
\begin{document}
\title{Conditional preparation of a quantum state in the continuous variable regime :\\
 generation of a sub-Poissonian state from twin beams}
\author{J. Laurat}
\author{ T. Coudreau}
\altaffiliation{A. Ma\^{\i}tre and T. Coudreau are also at the
P\^ole Mat{\'e}riaux et Ph{\'e}nom{\`e}nes Quantiques FR CNRS
2437, Universit{\'e} Denis Diderot, Paris}
\email{coudreau@spectro.jussieu.fr}
\author{N. Treps}
\author{A. Ma\^{\i}tre$^*$}
\author{C. Fabre } \affiliation{Laboratoire
Kastler Brossel, UPMC, Case 74, 4 Place Jussieu, 75252 Paris cedex
05, France}

\date{Received: 16 April 2003 / Revised version: 12 September 2003}

\begin{abstract}
We report the first experimental demonstration of conditional
preparation of a non-classical state of light in the continuous
variable regime. Starting from a non-degenerate OPO which
generates above threshold quantum intensity correlated signal and
idler "twin beams", we keep the recorded values of the signal
intensity only when the idler intensity falls inside a band of
values narrower than its standard deviation. By this very simple
technique, we generate a sub-Poissonian state 4.4 dB (64\%) below
shot noise from twin beams exhibiting 7.5 dB (82\%) of noise
reduction in the intensity difference.
\end{abstract}

 \maketitle

A well-known technique to generate a single photon state from
quantum correlated photons ("twin photons") is to use the method
of conditional measurement: if one labels (1) and (2) the two
modes in which the twin photons are emitted, it consists in
retaining in the information collected on mode (1) only the counts
occurring when a photon is detected in mode (2) (within a given
time window $\Delta T$). This method has been widely and very
successfully used over the past decades, firstly with twin photons
generated by an atomic cascade \cite{Aspect}, then by using the
more efficient technique of parametric down conversion
\cite{fock1}. Various protocols have been proposed to use
conditional preparation in order to generate other kinds of
non-classical states, for example Schr\"odinger cat states using a
squeezed vacuum state transmitted through a beamsplitter and a
measurement conditioned by the counts detected on the reflected
port \cite{cat}. In a similar way, teleportation of a quantum
state of light can be achieved by using conditional measurements
\cite{teleport} and the degree of entanglement can be improved by
photon subtractions \cite{improve}. In cavity QED, conditional
measurements on the atomic state have also led to the experimental
generation of non-classical photon states \cite{haroche}.

State reduction is obviously not restricted to the case of photon
counting, so that it may be interesting to extend this technique
to the continuous variable regime, where a continuously varying
photocurrent is measured instead of a series of photocounts.
Continuous detection conditioned by a photon counting event has
been implemented in various schemes \cite{OrozcoSchiller}. Closer
to our proposal where continuous measurements are used both for
triggering and characterizing the generated state, many
theoretical protocols have been suggested relying on two-mode
squeezed vacuum produced by a non-degenerate optical parametric
amplifier \cite{reducetwins,kumar}. For instance, homodyne
measurements on the idler can be used to condition the detection
of the signal and would reduce it to a squeezed state
\cite{reducetwins}.

In a conditional state preparation, the generation of the
non-classical state can be seen as the collapse of the entangled
wave-function induced by the measurement on one of its components.
However, very frequently, the measurement is made by
post-selection of the relevant events in the record of all the
values measured on the two channels, which can be made after the
end of the physical measurement, so that no wave-function collapse
actually occurs during the experiment. Note that, in contrast to
the methods of direct generation of a non-classical state, the
exact time window when the state is produced is not controlled in
a conditional measurement, and what we will call the "success
rate", i.e. the probability of generating the non-classical state
in a given time interval, is an important parameter to
characterize its efficiency.

To the best of our knowledge, no scheme has been suggested so far
to generate non-classical intense beams by the technique of
conditional measurement performed on continuous variables.  This
is the purpose of the present letter, which proposes a very simple
way of conditionally preparing a bright sub-Poissonian beam from
twin beams, and reports on its experimental implementation. The
theory of the presented technique will be detailed in a
forthcoming publication \cite{theory}.

It is well known that a non-degenerate optical parametric
oscillator (ND-OPO) produces above threshold intense signal and
idler "twin beams" \cite{traditional} : in the ideal case of a
system without losses, the Fourier components of the signal and
idler intensity quantum fluctuations which lie inside the cavity
bandwidth are perfectly correlated. The correlations are
characterized by the "gemellity", which is the remaining noise on
the intensity difference between the signal and idler intensities
normalized to the corresponding quantum noise level
\cite{Reynaud}. The instantaneous values of the signal and idler
photocurrents play therefore the role of the occurrence of counts
in the photon counting regime. The conditional measurement
technique that we propose consists in selecting the signal
photocurrent $I_s$ only during the time intervals when the idler
intensity $I_i$ has a given value $I_0$ (within a band $\Delta I$
smaller than the photocurrent standard deviation). The
measurements outside these time intervals are discarded. If the
correlation is perfect and the interval $\Delta I$ close to zero,
the recorded signal intensity is perfectly constant, and an
intense number state is generated by this conditional measurement.

In a real experiment, the correlation between the signal and idler
photocurrents is not perfect, and the selection band $\Delta I$ is
finite, so that the method will not prepare a perfect number
state, but a sub-Poissonian state instead. The density matrix
describing the state of light which is produced by such a state
reduction technique can be determined within the approximation
that the signal and idler photon distributions are Gaussian
\cite{theory}. In the limit where $\Delta I$ is very small, one
finds that, whatever the initial intensity noise of the beams are,
the conditional measurement produces a sub-Poissonian signal beam,
characterized by a Fano factor equal to the conditional variance
of the intensity fluctuations of the signal beam knowing the
intensity fluctuations of the idler beam, which plays an important
role in the characterization of Quantum Non Demolition
measurements \cite{grangier}. In other words, in the limit of
large correlations, the intensity noise reduction on the
conditionally prepared state is equal to the twin beam noise
reduction minus 3dB.

Obviously, if $\Delta I$ is very small, the probability that the
idler intensity lies within the chosen band is also very small,
and the success rate of the non-classical state production by such
a conditional measurement is also very low. Computer simulations,
as well as analytical calculations \cite{theory}, show that the
Fano factor of the generated state remains almost constant in a
wide range of $\Delta I$ values, whereas the success rate of the
method increases quickly. It is only when $\Delta I$ reaches
values comparable to the shot noise standard deviation that the
post-selection process becomes less efficient, and the Fano factor
tends to its uncorrected value.

The present conditional measurement technique has strong analogies
with the method of active feed-forward correction of the signal
beam intensity, by opto-electronics techniques, using the
information obtained from the measurement of the idler intensity
\cite{mertz1, mertz2}, which produces a sub-Poissonian state with
a Fano factor also equal to the conditional variance. The present
technique is much simpler to implement, whereas the active
correction technique is non-conditional, and has the advantage of
producing the non-classical state at all times.

\begin{figure}
\includegraphics[width=.9\columnwidth]{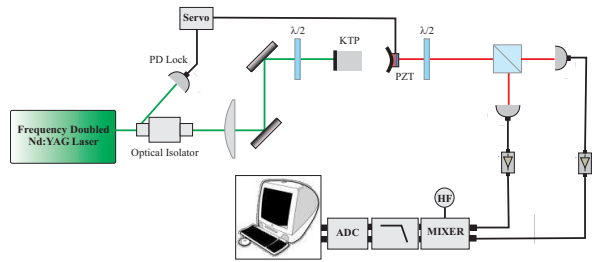}
\caption{\label{setup}Experimental layout. The 1064nm orthogonally
polarized bright twin beams generated by the ND-OPO are separated
by a polarizing beam splitter (PBS) and detected. The high
frequency components of the photocurrents are amplified and
demodulated at a given Fourier frequency. The two channels are
digitized and simultaneously recorded on a computer.}
\end{figure}

The experimental setup is shown in Figure \ref{setup}. A
continuous frequency-doubled Nd:YAG laser pumps a triply resonant
ND-OPO above threshold, made of a semi-monolithic linear cavity :
in order to increase the mechanical stability and reduce the
reflection losses, the input flat mirror is directly coated on one
face of the 10mm-long KTP crystal. The reflectivities for the
input coupler are 95.5\% for the pump (532nm) and almost 100\% for
the signal and idler beams (1064nm). The output coupler (R=38mm)
is highly reflecting for the pump and its transmission is 5\% for
the infrared. At exact triple resonance, the oscillation threshold
is less than 15mW. The OPO is actively locked on the pump
resonance by the Pound-Drever-Hall technique: we detect by
reflection the remaining 12MHz modulation used in the laser to
lock the external doubling cavity. In order to stabilize the OPO
infrared output intensity, the crystal temperature has to be
drastically controlled (within a mK). The OPO can operate stably
during more than one hour without mode-hopping. The signal and
idler orthogonally polarized beams (in the 1-5mW range) generated
by the OPO are then separated by a polarizing beam-splitter and
detected on a pair of balanced high quantum efficiency InGaAs
photodiodes (Epitaxx ETX300, quantum efficiency: 95\%). An
half-wave plate is inserted before the polarizing beam-splitter.
When the polarization of the twin beams is turned by 45$^\circ$
with respect to its axes, it behaves as a 50\% usual
beam-splitter, which allows us to measure the shot noise
level\cite{traditional}.

In almost all the bright twin beams experiments to date
\cite{traditional}, the photocurrents are subtracted and the
difference is sent onto a spectrum analyzer which gives the
variance of the photocurrent distribution. We have implemented a
different protocol to have access to the full photon-number
quantum statistics of the signal and idler beams at a given
Fourier frequency $\Omega$ (see also \cite{computer}): each
photocurrent is amplified and multiplied by a sinusoidal current
at frequency $\Omega$ produced by a signal generator, and filtered
by a 22kHz low-pass filter in order to obtain the instantaneous
value of the photocurrent Fourier component at frequency $\Omega$,
which is then digitized at a sampling rate of 200 kHz by a 12-bit,
4-channel acquisition card (National Instruments PCI-6110E), which
also simultaneously records the instantaneous values of the DC
photocurrents. Two successive acquisitions (200 000 points for
each channel) are required, one for calibrating the shot noise by
rotating the half-wave plate and the other to record the quantum
correlated signals.

\begin{figure}
\includegraphics[width=.9\columnwidth]{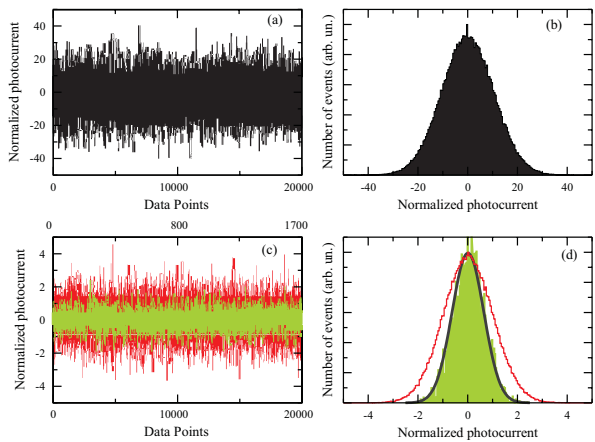}
\caption{\label{courbes_exp}Experimental results: (a) Idler
intensity fluctuations: 200 000 acquired points at 3.5 MHz
demodulation frequency (only 20 000 shown). (b) Corresponding
probability distribution. The unit is the width $\sigma_0$ of the
Poisson distribution of same mean intensity (shot noise). (c)
Values of the signal intensity conditionally selected by the value
of the idler intensity recorded at the same time (the selection
bandwidth $\Delta I$ equal to 0.1 $\sigma_0$ around the mean),
superimposed to the corresponding experimentally measured shot
noise. (d) Corresponding probability distribution, compared to the
Poisson distribution (grey line), displaying the sub-Poissonian
character of the conditionally generated state. The black line is
a gaussian fit of the intensity distribution}
\end{figure}

Figure \ref{courbes_exp} sums up the measurements obtained with a
demodulation frequency $\Omega /2 \pi =3.5 MHz$. Figure
\ref{courbes_exp}a shows the actual recording of the fluctuations
of the idler beam during a time interval of 100ms. As the OPO is
pumped close to threshold, the signal and idler beams have
intensity fluctuations which are much larger than the standard
quantum limit, as can be seen on curve \ref{courbes_exp}b, giving
the probability distribution of the intensity fluctuations
normalized to the shot noise. The corresponding Fano factor
exceeds 100 (20~dB above the shot noise level). From the recorded
data, one can calculate the noise variance on the difference
between the signal and idler intensities.  It reaches a value of
7.5~dB below the standard quantum limit (8.5~dB after correction
of dark noise), in good agreement with the value of the noise
variance measured on the spectrum analyzer. The dark noise, which
is more than 6~dB below all measurements, is not subtracted in the
following experimental results.

The conditional measurement is performed in the following way :
the signal intensity values are kept only if the idler intensity
values recorded at the same time fall inside a narrow band around
its mean value. The remaining ensemble of values of the signal
intensity is given in figure \ref{courbes_exp}c, in which the shot
noise is given at the same time : one indeed observes a
significant narrowing of the probability distribution below the
shot noise level. Let us stress that the shot noise is unchanged
by this selection process: the beam exists only in the selection
intervals so that the unselected intervals do not contribute to
the average value. Figure \ref{courbes_exp}d gives the probability
distribution of the intensity fluctuations of the conditionally
prepared state normalized to the shot noise, together with the
Poissonian distribution of photons for the same mean intensity.
With a selection bandwidth $\Delta I$ equal to 0.1 times the
standard deviation $\sigma_0$ of a coherent state having the same
power (shot noise level), the conditionally prepared light state
exhibits 4.4 dB of noise reduction below the Poisson distribution.
This value is very close from the theoretical expectation in the
case of vanishingly narrow intensity band. The success rate of the
conditional preparation is around 0.85\% (1700 points out of 200
000 are accepted). This value would be higher for an initial state
with less excess noise above shot noise.

\begin{figure}
\includegraphics[width=.9\columnwidth]{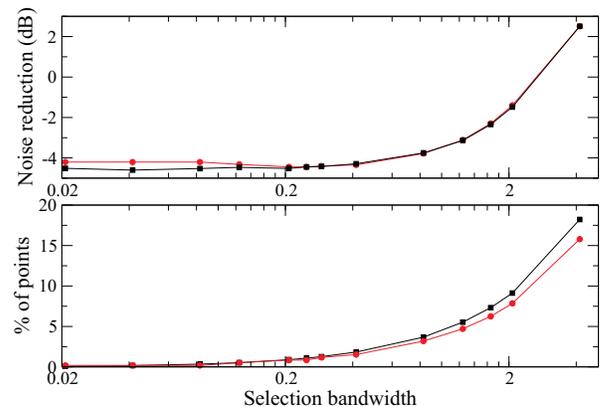}
\caption{\label{limite}Measured intensity noise on the
post-selected signal (a) and success rate (proportion of selected
points)(b) as a function of the selection bandwidth on the idler
normalized to $\sigma_0$. Circles: experimental data. Squares:
theoretical predictions.}
\end{figure}

The success rate can be improved by increasing the selection
bandwidth, at the expense of a decreased non-classical character
of the selected state. Figure \ref{limite} shows the measured
residual noise in the conditionally produced state, and the
success rate of the state generation, as a function of the
selection bandwidth normalized to the $\sigma_0$. The noise
reduction turns out to be almost constant until the normalized
selection bandwidth becomes of the order of 0.1, whereas the
success rate steadily increases, in very good agreement with
numerical simulations. However, one can see a slight increase in
the noise when the selection bandwidth becomes very narrow. This
artefact is due to the sampling process on a finite range of bits
(12), which restricts the resolution of the acquisition.

\begin{figure}
\includegraphics[width=.9\columnwidth]{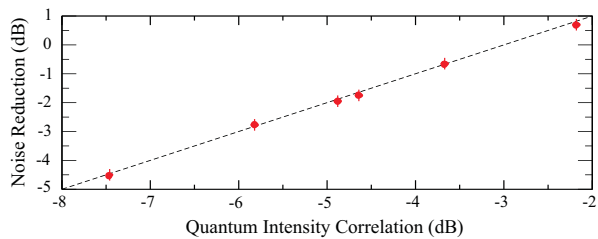}
\caption{\label{pertes}Measured intensity noise reduction for
different values of the signal-idler correlation. The selection
bandwidth is taken equal to 0.1 times the shot noise. Circles:
experimental data. Dotted line: theoretical prediction at the
limit of a very small selection bandwidth (gemellity minus 3 dB).}
\end{figure}

In figure \ref{pertes}, we give the measured residual noise for
different amounts of intensity correlations between the beams,
which can be varied by inserting losses on the OPO beams. One
checks on the figure the validity of the prediction that the noise
reduction is equal to the gemellity minus 3dB. One also observes
than, when the intensity difference noise is reduced by less than
3dB below the standard quantum level, the conditional state has
reduced intensity noise fluctuations in comparison with the very
noisy initial beam, but that it is not a non-classical
sub-Poissonian state.

\begin{figure}[!htpb]
\includegraphics[width=.9\columnwidth]{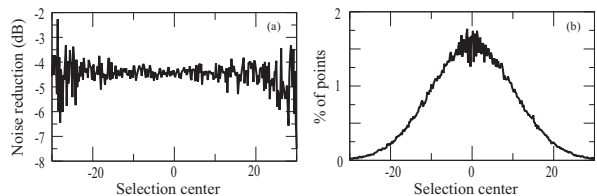}
\caption{\label{multiband} Measured intensity noise on the reduced
state (a) and success rate (b) as a function of the band center
normalized to $\sigma_0$. The selection bandwidth is taken equal
to 0.1 times the shot noise.}
\end{figure}

The continuous variable regime offers a unique possibility to
improve dramatically the efficiency of conditional strategy, by
choosing multiple selection bands with different mean values on
the idler intensity. Independent selection bands will correspond
to independent sets of time windows. By using hundreds of
independent intervals, one keeps most of the values of the signal,
each of these intervals reducing the signal to a given
sub-Poissonian state. Figure \ref{multiband} shows that the noise
reduction does not depend of the band center and that the success
rate follows the initial gaussian noise distribution.

To conclude, we have shown the first experimental demonstration of
conditional preparation of a quantum state in the continuous
variable regime. We have studied the influence of the selection
bandwidth of the conditioning measurement on the obtained
non-classical state and on the success rate of its preparation and
shown that many sub-Poissonian states can be produced in parallel.
This method to generate non-classical states of light in the
continuous variable regime is equivalent to sending the signal
beam through an intensity modulator which either totally transmits
the beam when the idler beam has the right value, or blocks it
when it is not the case. It therefore drastically changes the
light state, and seems to be very different from the usual
technique which consists of correcting the beam fluctuations by a
feed-forward or feed-back opto-electronic loop which only slightly
modifies the quantum fluctuations. This strongly non-linear
character of the action on the light may lead to the generation of
non-Gaussian states. One could also envision other criteria of
conditioning the quantum state than simply imposing to the idler
beam intensity to lie within a given band. This could also lead to
the generation of new families of non-classical bright states of
light.
\begin{acknowledgments}
Laboratoire Kastler-Brossel, of the Ecole Normale Sup\'{e}rieure
and the Universit\'{e} Pierre et Marie Curie, is associated with
the Centre National de la Recherche Scientifique (UMR 8552). We
acknowledge support from the European Commission project QUICOV
(IST-1999-13071) and ACI Photonique (Minist\`ere de la Recherche).
\end{acknowledgments}

\end{document}